\documentclass[%
 reprint,
 amsmath,amssymb,
 aps,
]{revtex4-2}

\usepackage{graphicx}% Include figure files
\usepackage{dcolumn}% Align table columns on decimal point
\usepackage{bm}% bold math
\usepackage{hyperref}
\usepackage{breakurl}

\usepackage{cleveref}
\newcommand{\crefrangeconjunction}{--}
\crefname{figure}{FIG.}{FIG.}
\crefname{equation}{}{}
\crefname{section}{}{}

\begin{document}

\preprint{APS/123-QED}

\title{Exceptional points in cylindrical elastic media with radiation loss}% Force line breaks with \\

\author{Kei Matsushima}
\email{matsushima@mech.t.u-tokyo.ac.jp}
\author{Yuki Noguchi}%
\author{Takayuki Yamada}%
\affiliation{%
Department of Mechanical Engineering, Graduate School of Engineering, the University of Tokyo
\\
2-11-16 Yayoi, Bunkyo-ku, Tokyo, Japan
}%

\date{\today}% It is always \today, today,

\begin{abstract}
Exceptional points (EPs) are singular points on a parameter space at which some eigenvalues (scattering poles) and their corresponding eigenmodes coalesce. This study shows the existence of second- and third-order EPs in cylindrical elastic systems with radiation loss. We consider multilayered cylindrical solids under the plane-strain condition placed in a background elastic or acoustic medium. Elastic and acoustic waves propagating in the background media are subject to the radiation loss. We optimize the radii and the material constants of the multilayered solids, such that some scattering poles coalesce on the complex frequency plane. Some numerical experiments are performed to confirm that the coalescence originates from EPs. We expect that this study provides a new approach for enhancing mechanical sensors.
\end{abstract}

\maketitle

%\tableofcontents

\section{Introduction}
% Introduction on EP
Exceptional points (EPs) are singularities in a parameter space at which both eigenvalues and eigenmodes coalesce \cite{kato1995perturbation,berry2004physics,heiss2012physics}. This simultaneous degeneracy is a unique phenomenon of non-Hermitian systems. Unlike Hermitian systems, non-Hermitian systems have complex eigenvalues, and their imaginary part represents the rate of exponential decay or divergence of resonances in time. Around an EP, two (or more) eigenvalues are associated with each other and draw a Riemann surface of a multivalued root function on the parameter space. This special structure can be observed by continuously changing the parameters around an EP and tracking the corresponding eigenvalues, called EP encircling \cite{dembowski2001experimental}. As the root-like behavior improves the sensitivity of the eigenvalue splitting, a promising application of the non-Hermitian degeneracy is the enhancement of optical and mechanical sensors \cite{wiersig2014enhancing,wiersig2016sensors,chen2017exceptional,hokmabadi2019nonhermitian,xing2020ultrahigh,yu2021whisperinggallerymode,yuan2022exceptional}.

% PT-symmetric systems
Parity-time (PT) symmetry is one of the most successful approaches for realizing EPs \cite{klaiman2008visualization,longhi2009bloch,ruter2010observation,ozdemir2019parity}. A system is said to be PT symmetric if the same amount of gain and loss is symmetrically distributed in space. Due to these gain and loss, PT-symmetric systems are non-Hermitian and possibly exhibit EPs. If a system is PT symmetric, the non-Hermitian degeneracy occurs at the PT phase transition points, which can be easily found by tuning the amount of gain and loss \cite{chong2011mathcalp}.

% non PT symmetric
Although the PT symmetry is an important property when discussing non-Hermitian physics, the EP itself is ubiquitous for various non-Hermitian systems. For example, dielectric microcavities can induce the non-Hermitian degeneracy without gain by tuning the dielectric constant and/or the system geometry \cite{cao2015dielectric}. Kullig et al. recently showed the existence of second- and third-order EPs in two-dimensional multilayered microdisk cavities \cite{kullig2018exceptional}. They optimized the radii and refractive indices of the layered cavities such that two or three whispering gallary modes coalesce at a single complex frequency. Bulgakov et al. proved that a spheroid exhibits a second-order EP \cite{bulgakov2021exceptional}. Abdrabou and Lu studied EPs that arise in two-dimensional multiple scattering systems \cite{abdrabou2019exceptional}. Meanwhile, Gwak et al. realized steady EPs by deforming circular microcavity \cite{gwak2021rayleigh}. Photonic crystal slabs also exhibit EPs without material gain \cite{zhen2015spawning,kaminski2017control,abdrabou2018exceptional,zhou2018observation,abdrabou2020exceptional}.

% Elastic systems
EPs can also be observed in acoustic and elastic systems. 
As mentioned, PT symmetry is a convenient platform for discussing non-Hermitian effects on acoustic and elastic wave propagation. Recent studies revealed that the well-known anomalies of PT symmetric systems (e.g., phase transition, asymmetrical reflectance, lasing/anti-lasing, negative refraction, and enhanced sensitivity) are observed in such systems \cite{zhu2014mathcalp,fleury2015invisible,fleury2016paritytime,shi2016accessing,christensen2016paritytime,achilleos2017nonhermitian,auregan2017mathcalp,hou2018mathcalp,wu2019asymmetric,wang2019extremely,dominguez-rocha2020environmentally,rosa2021exceptional,farhat2021selfdual,cai2022exceptional,yi2022structural}. However, the existence of EPs in two- and three-dimensional passive elastic systems is still uncertain, except for an open periodic system \cite{mokhtari2020scattering}.

% Objective and scope
In this study, we numerically show that second- and third-order EPs exist in cylindrical elastic media without gain. The system comprises a multilayered cylindrical solid and background elastic or acoustic medium. The background medium is unbounded in space and induces the energy loss due to the radiation in the radial direction. The solid--solid and solid--fluid coupled problems are solved via Helmholtz decomposition and cylindrical functions. Following \cite{kullig2018exceptional}, we optimize the radii and material constants such that two or three eigenvalues (scattering poles) coalesce on the complex frequency plane. Subsequently, the EP encircling is performed around the optimized parameters to confirm that the coalescence originates from the non-Hermitian degeneracy. Moreover, we show that the degenerate poles behave as multivalued root functions around EPs, which is an important property for enhancing mechanical sensors.

\section{Model}
\begin{figure}[b]
  \includegraphics[scale=0.4]{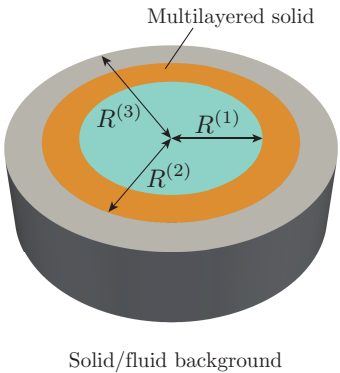}% Here is how to import EPS art
  \caption{\label{fig:layer} Multilayered elastic solid embedded in an elastic or acoustic medium. The figure illustrates the case of $N = 3$.}
\end{figure}

We consider the time-harmonic oscillation of a multilayered solid embedded in a background elastic or acoustic medium as shown in \cref{fig:layer}. Each solid layer, indexed by the integer $l=1,\ldots,N$ (from innermost to outermost), is a linear, homogeneous, and isotropic elastic medium with mass density $\rho^{(l)}$ and Lam\'e constants $(\lambda^{(l)},\mu^{(l)})$. We also assume that the solid is under the plane-strain condition. The in-plane displacement $u$ and stress $\sigma$ are subject to the following two-dimensional Navier--Cauchy equations:
\begin{align}
  \nabla \cdot \sigma + \rho^{(l)}\omega^2 u = 0 , \label{eq:navier}
\end{align}
where $\omega$ is the angular frequency. The stress tensor $\sigma$ is associated with the displacement $u$ via the following linear relations:
\begin{align}
  \sigma_{ij} &= \lambda^{(l)} (\nabla \cdot u) \delta_{ij} + 2\mu^{(l)} \varepsilon_{ij},
  \\
  \varepsilon_{ij} &= \frac{1}{2}\left( \frac{\partial u_i}{\partial x_j} + \frac{\partial u_j}{\partial x_i} \right),
\end{align}
where $\delta_{ij}$ is the Kronecker delta.
The displacement $u$ and the normal component $\sigma\cdot n$ of the stress are continuous on the solid--solid interfaces, where $n$ is the unit normal vector.

In what follows, we assume that the solid is a cylinder, i.e., the cross section $\Omega^{(l)}$ of Layer $l$ is given by $\Omega^{(l)} = \{ x\in\mathbb{R}^2 \mid R^{(l-1)} < |x| < R^{(l)} \}$ with outer radius $R^{(l)}$ ($R^{(0)}=0$).

A solution of the Navier--Cauchy equations \cref{eq:navier} can be sought via Lam\'e potentials $\phi^{(l)}$ and $\psi^{(l)}$ defined as
\begin{align}
  u_i = \frac{\partial \phi^{(l)}}{\partial x_i} + e_{3ij} \frac{\partial \psi^{(l)}}{\partial x_j} \quad\text{in }\Omega^{(l)},
\end{align}
where $e_{ijk}$ is the Levi--Civita symbol \cite{eringen1975elastodynamics}. The potentials decompose the Navier--Cauchy equations \cref{eq:navier} into the following Helmholtz equations:
\begin{align}
  \nabla^2 \phi^{(l)} + \left({\omega /c_\mathrm{L}^{(l)}} \right)^2 \phi^{(l)} = 0 \quad\text{in }\Omega^{(l)},
  \\
  \nabla^2 \psi^{(l)} + \left({\omega /c_\mathrm{T}^{(l)}} \right)^2 \psi^{(l)} = 0 \quad\text{in }\Omega^{(l)},
\end{align}
where $c_\mathrm{L}^{(l)} = \sqrt{(\lambda^{(l)}+2\mu^{(l)})/\rho^{(l)}}$ and $c_\mathrm{T}^{(l)}= \sqrt{\mu^{(l)}/\rho^{(l)}}$ are the phase speed of the longitudinal and transverse waves in $\Omega^{(l)}$, respectively. We can write a solution $(\phi^{(l)},\psi^{(l)})$ as follows for each azimuthal index $m\in\mathbb{Z}$:
\begin{align}
  \phi^{(l)} &= A_\mathrm{L}^{(l)} J_m(\omega r/c_\mathrm{L}^{(l)}) \exp(\mathrm{i} m \theta)
  \notag\\
  &+ B_\mathrm{L}^{(l)} H^{(1)}_m(\omega r/c_\mathrm{L}^{(l)}) \exp(\mathrm{i} m \theta),
  \\
  \psi^{(l)} &= A_\mathrm{T}^{(l)} J_m(\omega r/c_\mathrm{T}^{(l)}) \exp(\mathrm{i} m \theta)
  \notag\\
  &+ B_\mathrm{T}^{(l)} H^{(1)}_m(\omega r/c_\mathrm{T}^{(l)}) \exp(\mathrm{i} m \theta),
\end{align}
where $(r,\theta)$ is the polar coordinate system. For $l=1,\ldots,N-1$, the continuity at the interface $r=R^{(l)}$ associates the unknown coefficients $A^{(l)}_\mathrm{L}$, $A^{(l)}_\mathrm{T}$, $B^{(l)}_\mathrm{L}$, and $B^{(l)}_\mathrm{T}$ as
\begin{align}
  M^{(l+1)}(R^{(l)})
  \begin{pmatrix}
    A_\mathrm{L}^{(l+1)} \\ A_\mathrm{T}^{(l+1)} \\ B_\mathrm{L}^{(l+1)} \\ B_\mathrm{T}^{(l+1)}
  \end{pmatrix}
  =
  M^{(l)}(R^{(l)})
  \begin{pmatrix}
    A_\mathrm{L}^{(l)} \\ A_\mathrm{T}^{(l)} \\ B_\mathrm{L}^{(l)} \\ B_\mathrm{T}^{(l)}
  \end{pmatrix}
  , \label{eq:among_layers}
\end{align}
where the matrix $M^{(l)}(R)\in\mathbb{C}^{4\times 4}$ is dependent on the frequency $\omega$, azimuthal index $m$, and material constants $\rho^{(l)},\lambda^{(l)},\mu^{(l)}$. The explicit form of $M^{(l)}(R)$ is presented in Appendix A. Any solution should be bounded at $r=0$; thus, we have $B_\mathrm{L}^{(1)}=B_\mathrm{T}^{(1)}=0$. Accordingly, we can define a 4-by-2 matrix $X$ that gives
\begin{align}
  X
  \begin{pmatrix}
    A_\mathrm{L}^{(1)} \\ A_\mathrm{T}^{(1)}
  \end{pmatrix}
  =
  \begin{pmatrix}
    A_\mathrm{L}^{(N)} \\ A_\mathrm{T}^{(N)} \\ B_\mathrm{L}^{(N)} \\ B_\mathrm{T}^{(N)}
  \end{pmatrix}
  . \label{eq:among_layers_2}
\end{align}

\subsection{Solid background}
First, we formulate a radiation behavior in the background $r>R^{(N)}$. When the background medium is an elastic medium with a mass density $\rho$ and Lam\'e constants $(\lambda,\mu)$, we define two potentials $\phi$ and $\psi$ as follows in an analogous manner:
\begin{align}
  u_i = \frac{\partial \phi}{\partial x_i} + e_{3ij} \frac{\partial \psi}{\partial x_j} \quad r > R^{(N)}.
\end{align}
We are interested in a radiating field without incident waves; therefore, the potentials $\phi$ and $\psi$ are written as
\begin{align}
  \phi &= 
  B_\mathrm{L} H^{(1)}_m(\omega r/c_\mathrm{L}) \exp(\mathrm{i} m \theta),
  \\
  \psi &= 
  B_\mathrm{T} H^{(1)}_m(\omega r/c_\mathrm{T}) \exp(\mathrm{i} m \theta),
\end{align}
where $B_\mathrm{L}$ and $B_\mathrm{T}$ are constants, and $c_\mathrm{L}$ and $c_\mathrm{T}$ are defined by $c_\mathrm{L} = \sqrt{(\lambda+2\mu)/\rho}$ and $c_\mathrm{T}= \sqrt{\mu/\rho}$, respectively. Similar to Eq. \cref{eq:among_layers}, the continuity conditions at $r=R^{(N)}$ equate the coefficients as
\begin{align}
  Y_\mathrm{s}
  \begin{pmatrix}
    B_\mathrm{L} \\ B_\mathrm{T}
  \end{pmatrix}
  =
  Z_\mathrm{s}
  \begin{pmatrix}
    A_\mathrm{L}^{(N)} \\ A_\mathrm{T}^{(N)} \\ B_\mathrm{L}^{(N)} \\ B_\mathrm{T}^{(N)}
  \end{pmatrix}
  , \label{eq:solid-solid}
\end{align}
where the 4-by-2 matrix $Y_\mathrm{s}$ is defined in Appendix A, and $Z_\mathrm{s}=M^{(N)}(R^{(N)})$. Equations \cref{eq:among_layers_2,eq:solid-solid} give the following linear equations:
\begin{align}
  \begin{bmatrix}
    Z_\mathrm{s}X & -Y_\mathrm{s}
  \end{bmatrix}
  \begin{pmatrix}
    A^{(1)}_\mathrm{L} \\ A^{(1)}_\mathrm{T} \\ B_\mathrm{L} \\ B_\mathrm{T}
  \end{pmatrix}
  =
  \begin{pmatrix}
    0 \\ 0 \\ 0 \\ 0
  \end{pmatrix}
  . 
  \label{eq:solid}
\end{align}

\subsection{Fluid background}
The fluid background can be modeled in a similar fashion. Let $p$ be a pressure field that satisfies
\begin{align}
  \nabla^2 p + (\omega/c)^2 p = 0, \quad r>R^{(N)},
\end{align}
where the sound speed $c$ is given by the mass density $\rho$ and bulk modulus $\kappa$ of the background fluid as $c=\sqrt{\kappa/\rho}$. In analogy with the elastic case, the pressure $p$ admits the following solution with a constant $B$:
\begin{align}
  p = B H^{(1)}_m(\omega r/c) \exp(\mathrm{i} m \theta), \quad r>R^{(N)}.
\end{align}

Elastic--acoustic coupling is established via the continuity of the displacement and traction at $r=R^{(N)}$:
\begin{align}
  \sigma\cdot n = -pn,
  \\
  u\cdot n = \frac{1}{\rho\omega^2}\nabla p \cdot n.
\end{align}
Using the polar coordinate system, these conditions are translated into
\begin{align}
  Y_\mathrm{f} B = Z_\mathrm{f}
  \begin{pmatrix}
    A_\mathrm{L}^{(N)} \\ A_\mathrm{T}^{(N)} \\ B_\mathrm{L}^{(N)} \\ B_\mathrm{T}^{(N)}
  \end{pmatrix}
  , \label{eq:solid-fluid}
\end{align}
where the 3-by-1 matrix $Y_\mathrm{f}$ and 3-by-4 matrix $Z_\mathrm{f}$ are given in Appendix A.

Finally, we use Eqs. \cref{eq:among_layers_2,eq:solid-fluid} to obtain
\begin{align}
  \begin{bmatrix}
    Z_\mathrm{f} X & -Y_\mathrm{f}
  \end{bmatrix}
  \begin{pmatrix}
    A^{(1)}_\mathrm{L} \\ A^{(1)}_\mathrm{T} \\ B
  \end{pmatrix}
  =
  \begin{pmatrix}
    0 \\ 0 \\ 0
  \end{pmatrix}
  .
  \label{eq:fluid}
\end{align}

\section{Exceptional points}
Without any excitation, the time-harmonic oscillation of the multilayered solid is characterized by either Eq. \cref{eq:solid} or \cref{eq:fluid}. These linear systems always have a  trivial solution (i.e., uniformly $u=0$) and possibly nontrivial solutions for discrete angular frequencies $\omega\in\mathbb{C}$ when the determinant of the coefficient matrices becomes zero. These (angular) frequencies are often called {quasinormal mode frequencies}, {complex eigenvalues}, or {scattering poles}. The imaginary part of a scattering pole represents the reciprocal of the lifetime of the corresponding resonance, i.e., exponential decay rate in time induced by the radiation.

The scattering poles are dependent on the system parameters, e.g., the radii and material constants of the layered structure. When two (or more) scattering poles and corresponding resonant modes simultaneously coalesce on the complex $\omega$-plane for certain system parameters, the parameters are called an {exceptional point} on the parameter space.

In this study, we show that the multilayered elastic systems exhibit EPs using an optimization algorithm.

To prove that the systems exhibit an EP, we first numerically seek a nontrivial solution of the linear systems \cref{eq:solid} and \cref{eq:fluid}. The coefficient matrices depend on $\omega$ nonlinearly; hence, this is a nonlinear eigenvalue problem in terms of $\omega$. We solve this nonlinear eigenvalue problem using the Sakurai--Sugiura method \cite{asakura2009numerical}. Accordingly, we formally write the nonlinear eigenvalue problem as $\mathcal{A}(\omega)\xi = 0$, where $\mathcal{A}$ is a matrix-valued function. The Sakurai--Sugiura method converts the nonlinear eigenvalue problem into a linear one using contour integrals $\int_\gamma W^H \mathcal{A}^{-1}(\omega) V \mathrm{d}\omega$ for some full-rank matrices $W$ and $V$. The linear problem is then solved by a standard eigenvalue solver, which yields the complex eigenvalues (matrix poles) of $\mathcal{A}$ inside the contour path $\gamma$ and corresponding right eigenvector $\xi$. For more details, see \cite{asakura2009numerical}.

Given $M$ scattering poles $\omega_i$ ($i=1,\ldots,M$), we solve the following optimization problem:
\begin{align}
  \min_{\zeta\in\mathcal{P}} J(\zeta) = \sum_{i\neq j} |\omega_i(\zeta) - \omega_j(\zeta)|^2,
\end{align}
where $\mathcal{P}$ denotes a parameter space. A solution $\zeta$ is an EP of order $M$ in $\mathcal{P}$ if the optimization attains $J=0$ (i.e., $\omega_1=\ldots=\omega_M$), and all the corresponding resonant modes are identical.
%If the optimization attains $J=0$ (i.e., $\omega_1=\ldots=\omega_M$) and all the corresponding resonant modes $u_i$ are identical, we can say that the solution $\zeta$ is an exceptional point of order $M$ in $\mathcal{P}$.

\section{Results}
\subsection{Solid background}
In this subsection, we will show that the solid--solid system exhibits second- and third-order EPs. First, we consider the double-layered solid ($N=2$) embedded in a background medium with $(\rho,\lambda,\mu)$. In analogy with optical layered systems \cite{kullig2018exceptional}, we expect that non-Hermitian degeneracy is achieved when two whispering gallary modes coalesce at the same frequency. This can be accomplished when the refractive index of $\Omega^{(1)}$ is larger than that of $\Omega^{(2)}$. In what follows, we limit the consideration to the case of $m=\pm 8$. Poisson's ratio $\nu^{(l)}$ is uniformly $0.3$ for all elastic media.

\begin{figure*}
  \includegraphics[scale=0.35]{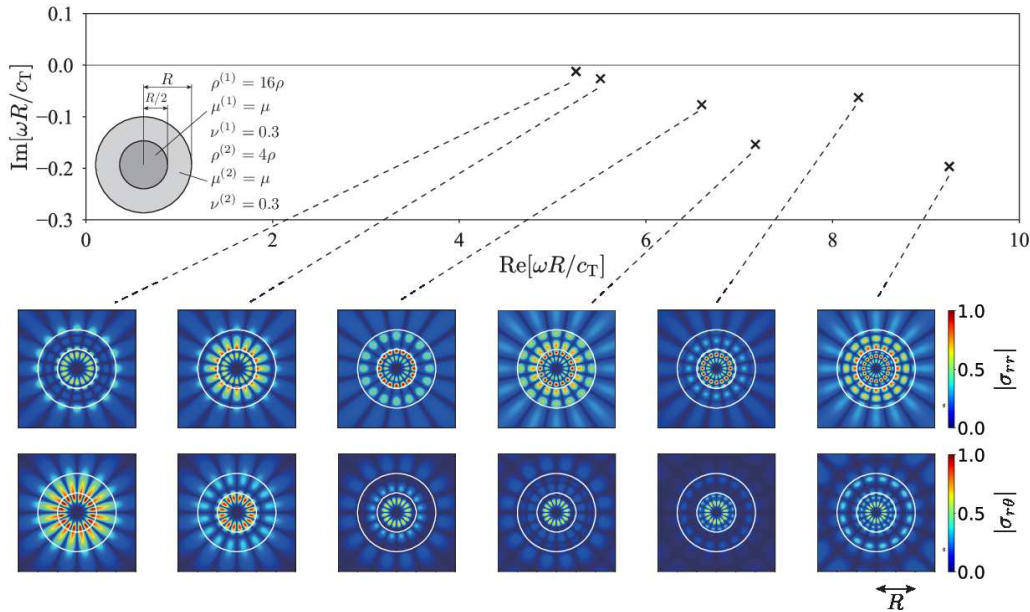}% Here is how to import EPS art
  \caption{\label{fig:solid-solid_EP2_init} Scattering poles and corresponding resonant modes of the double layers in the solid background.}
\end{figure*}
\cref{fig:solid-solid_EP2_init} shows the distribution of the scattering poles and corresponding resonant modes for $\rho^{(1)} = 16\rho$, $\rho^{(2)}=4\rho$, $\mu^{(1)}=\mu^{(2)}=\mu$, and $R^{(2)} = 2R^{(1)}=:R$. The result clearly indicates that the multilayered elastic system exhibits whispering gallary modes around the solid interfaces. We focus herein on the two neighboring scattering poles, $\omega_1 R/c_\mathrm{T} = 5.2513 - 0.012416\mathrm{i}$ and $\omega_2 R/c_\mathrm{T} = 5.5130 - 0.026387\mathrm{i}$, and seek a second-order EP on the $R^{(1)}$--$\rho^{(1)}$ parameter plane.

\begin{figure}
  \includegraphics[scale=0.2]{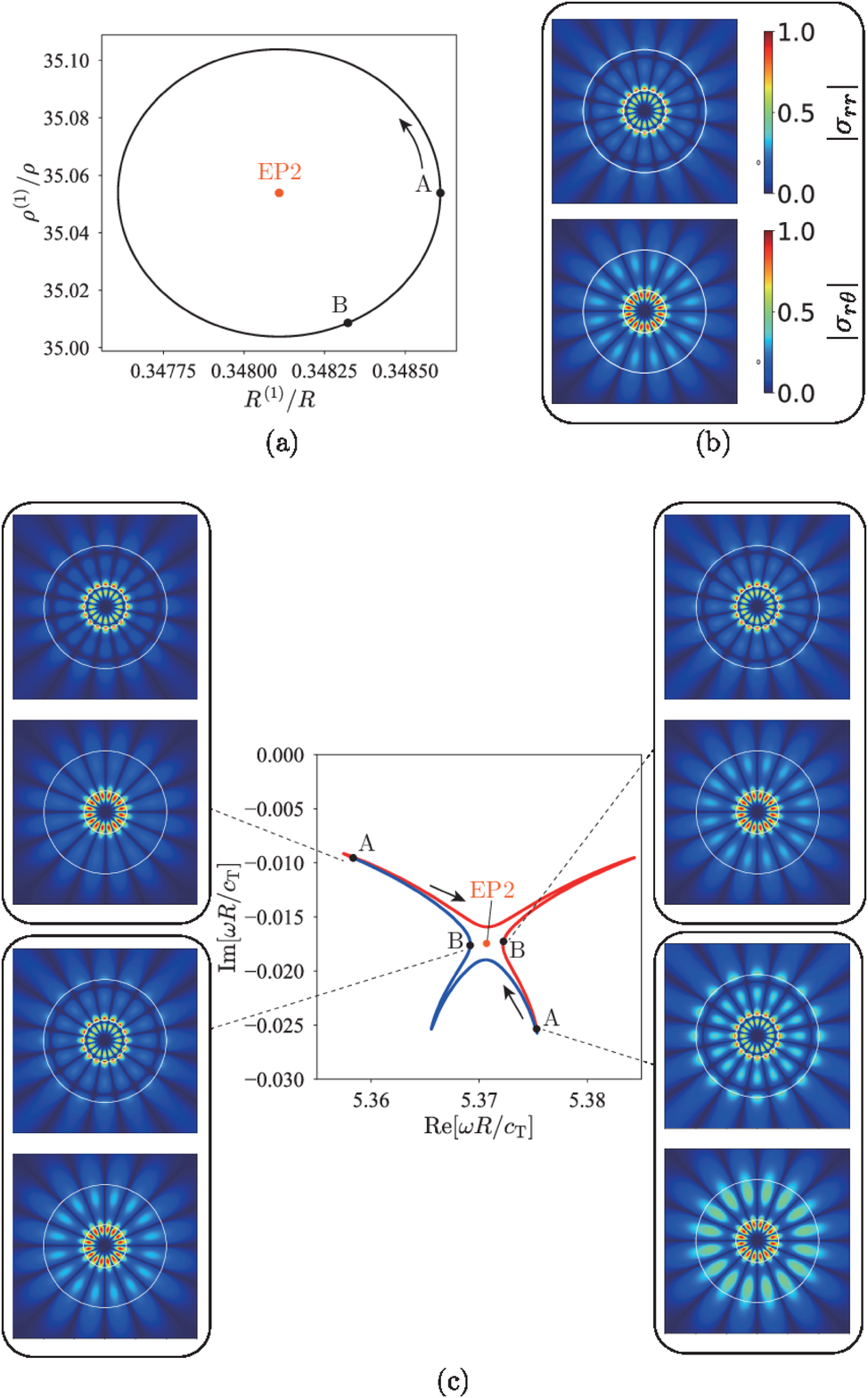}% Here is how to import EPS art
  \caption{\label{fig:solid-solid_EP2_result} Encircling a second-order EP (EP2) in the solid--solid system. (a) Encircling path in the parameter space. (b) Degenerate resonant mode at the EP2. (c) Trajectory of the two scattering poles. Each color corresponds to a single encircling shown in (a).}
\end{figure}
We performed a two-variable optimization using the Nelder--Mead method \cite{gao2012implementing} and found an optimal solution $R^{(1)}/R = 0.34811$ and $\rho^{(1)}/\rho = 35.054$. This optimized system has two scattering poles at $\omega R/c_\mathrm{T} = 
5.37067922-0.01744497\mathrm{i},
5.37067926-0.01744494\mathrm{i}$. The corresponding resonant modes, shown in \cref{fig:solid-solid_EP2_result} (b), also coalesce at the parameters. The simultaneous degeneracy of the poles and resonant modes is a proof of the existence of a second-order EP.
% 0.3481089659595145	35.053835577683714
% 5.37067922316748e+00_-1.74449712639752e-02.dat  5.37067926114971e+00_-1.74449378082571e-02.dat

\begin{figure}
  \includegraphics[scale=0.55]{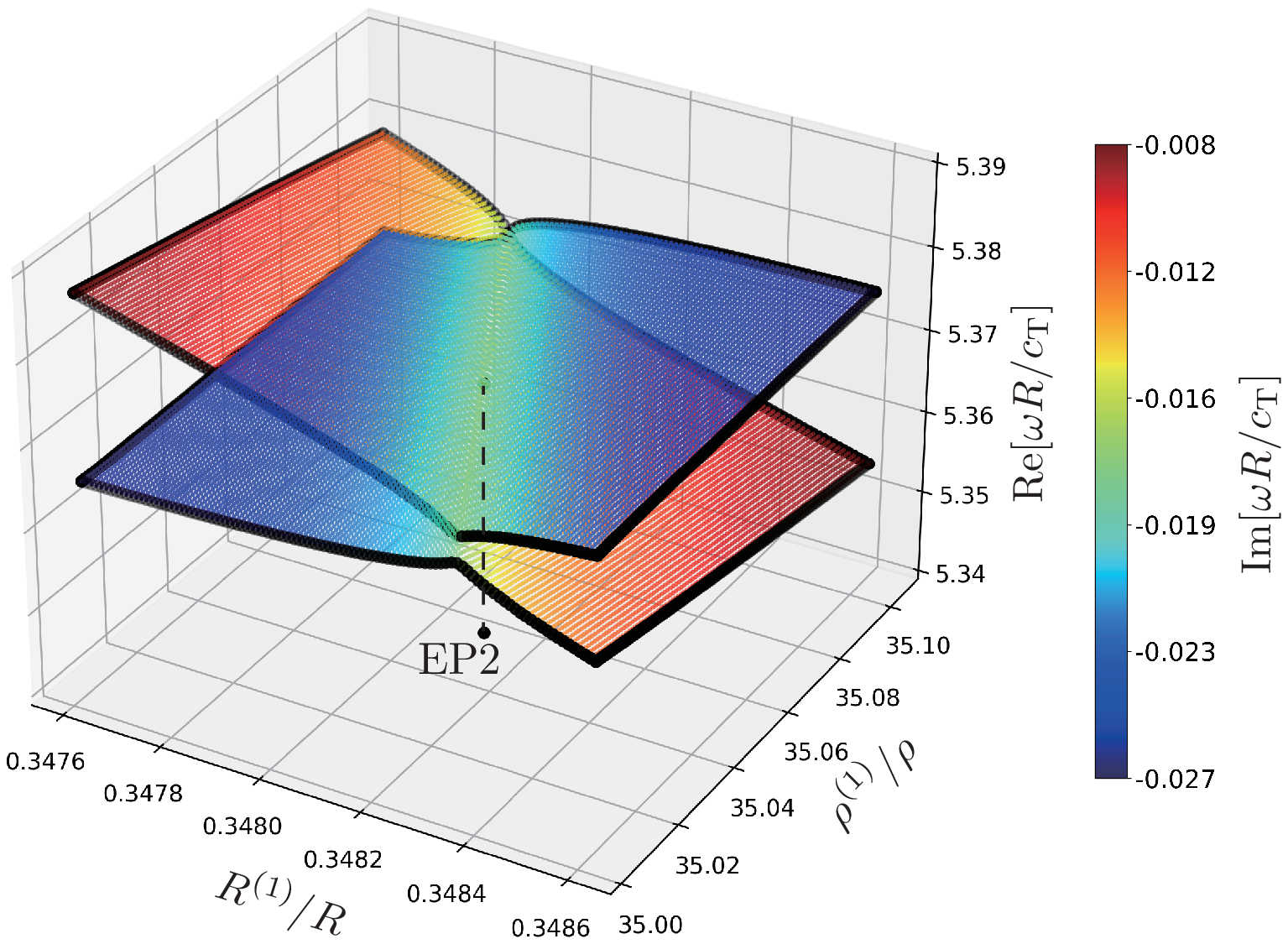}% Here is how to import EPS art
  \caption{\label{fig:solid-solid_EP2_sheet} Real and imaginary parts of the scattering poles for various parameters $R^{(1)}$ and $\rho^{(1)}$ around the EP2.}
\end{figure}
The unique characteristics of second-order EPs can be confirmed by encircling them on a two-variable parameter space. \cref{fig:solid-solid_EP2_result} (a) defines a closed elliptic path on the $R^{(1)}$--$\rho^{(1)}$ plane. The center of the path is the obtained EP. By tracing the path counterclockwise, we compute the two scattering poles on the complex $\omega$-plane and plot their trajectories in \cref{fig:solid-solid_EP2_result} (c). The results show that a single pole does not form a closed path on the $\omega$-plane although the parameter curve does on the $R^{(1)}$--$\rho^{(1)}$ plane. Instead, the pair of the two poles closes the trajectory. Equivalently, a single pole forms a closed loop when the parameter curve encircles the EP twice. This feature is analogous to the multivalued square root function on the complex plane.

We discuss this square root-like behavior by computing the pole pair for various parameters $R^{(1)}$ and $\rho^{(1)}$ around the EP. \cref{fig:solid-solid_EP2_sheet} illustrates the result. The pair forms two crossed sheets that branch off at the EP. This structure is similar to the Riemann surfaces of the square root function. These results imply that the obtained EP originates from a branch point of a complex-valued function, although it is not explicit in the equation \cref{eq:solid}.

%
% EP3
%
\begin{figure}
  \includegraphics[scale=0.2]{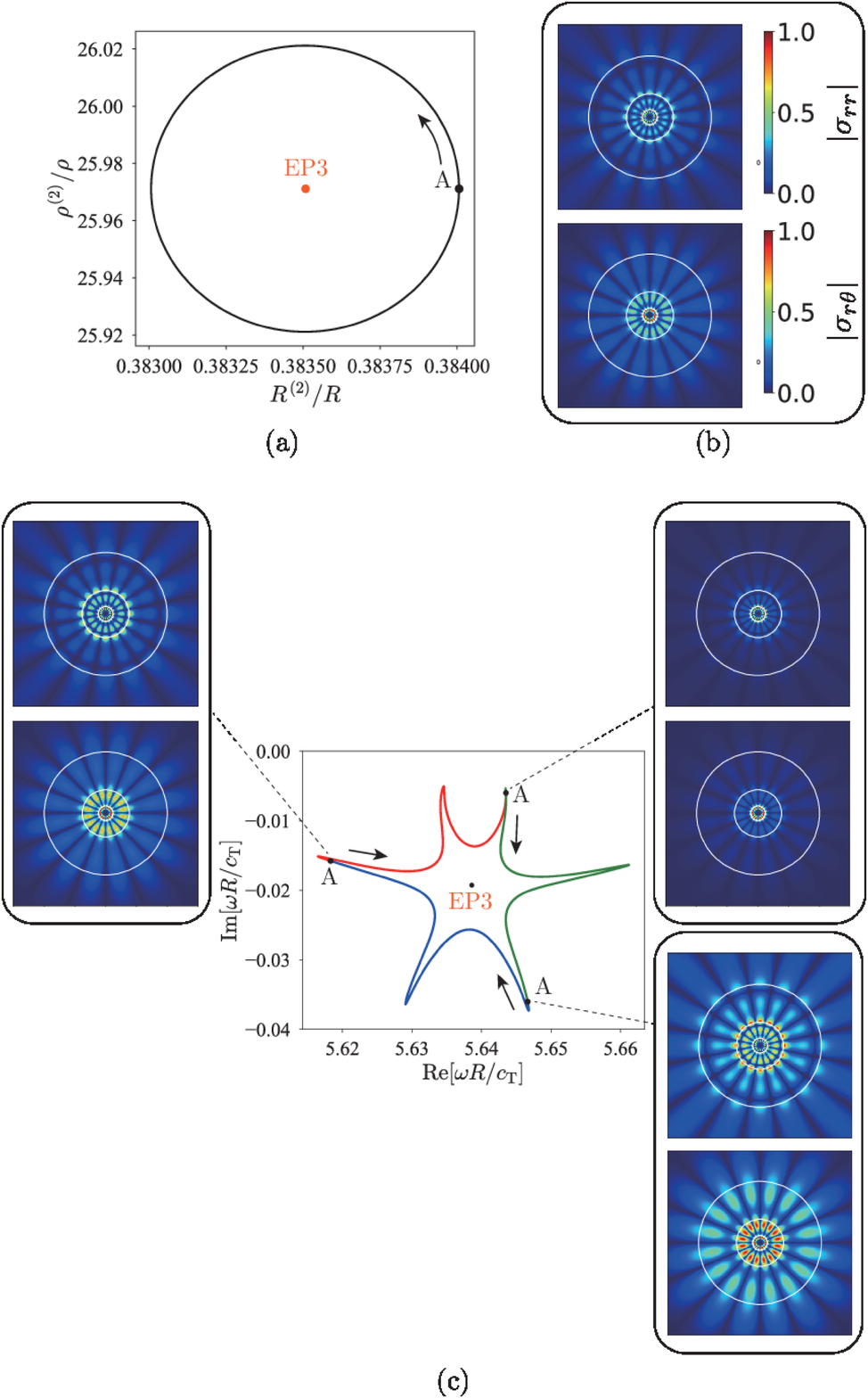}% Here is how to import EPS art
  \caption{\label{fig:solid-solid_EP3_result} Encircling a third-order EP (EP3) in the solid-solid system. (a) Encircling path in the parameter space. (b) Degenerate resonant mode at the EP3. (c) Trajectory of the two scattering poles. Each color corresponds to the single encircling shown in (a).}
\end{figure}
We now show that a triple-layered solid exhibits a third-order EP by starting with the following parameters:
\begin{itemize}
  \item Mass density $\rho^{(1)}=64\rho,\,\rho^{(2)}=16\rho,\,\rho^{(3)}=4\rho$
  \item Shear modulus $\mu^{(1)}=\mu^{(2)}=\mu^{(3)}=\mu$
  \item Radii $R^{(1)} = R/4$, $R^{(2)} = R/2$, $R^{(3)} =: R$.
\end{itemize}
We chose three scattering poles $\omega R/c_\mathrm{T} = 5.2088 - 0.005504\mathrm{i}$, $5.3657 - 0.016873\mathrm{i}$, and $5.5851 - 0.017023\mathrm{i}$ and optimized the five parameters $R^{(1)}$, $R^{(2)}$, $\rho^{(1)}$, $\rho^{(2)}$, and $\rho^{(3)}$ to find an EP.
%performed the objective function $J$.
%w1 = 5.208822491769 -0.005503709830325997j
%w2 = 5.365687571739271 -0.016873304515178475j
%w3 = 5.585084613364654 -0.017023089847521274j

The optimized parameters are $R^{(1)}/R=0.11969$, $R^{(2)}=0.38351$, $\rho^{(1)}/\rho = 270.69$, $\rho^{(2)}/\rho=25.971$, and $\rho^{(3)}/\rho=3.57568$. The optimized system has three poles $\omega R/c_\mathrm{T}$ at $5.6385 - 0.019254\mathrm{i}$, $5.6387 - 0.019293\mathrm{i}$, and $5.6386 - 0.019242\mathrm{i}$, which are almost degenerated. Moreover, \cref{fig:solid-solid_EP3_result} (b) show that not only the pole values but also corresponding resonant modes coalesced at the point. We confirm that the degeneracy arose from a third-order EP by performing the EP encircling on the $R^{(2)}$--$\rho^{(2)}$ parameter space. Namely, we continuously change the values of $R^{(2)}$ and $\rho^{(2)}$ and calculate the poles on the complex $\omega$-plane while the other parameters remain fixed. 

\begin{figure}
  \includegraphics[scale=0.85]{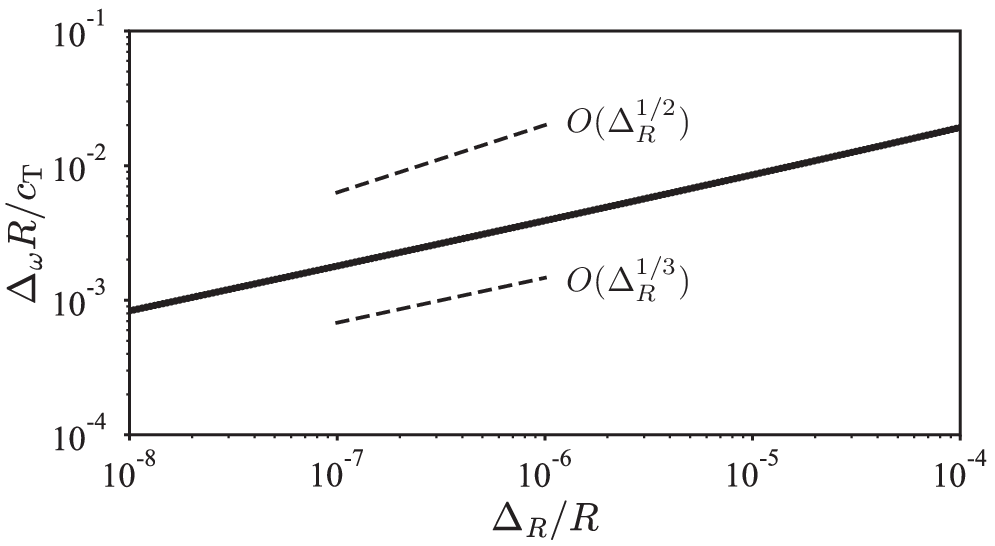}% Here is how to import EPS art
  \caption{\label{fig:solid-solid_EP3_distance} Variation of the degenerate poles at the third-order EP when the parameter $R^{(2)}$ is perturbed. }
\end{figure}

\cref{fig:solid-solid_EP3_result} depicts the results. As with the previous discussion on the second-order EP, we move the two parameters counterclockwise along the elliptic path shown in \cref{fig:solid-solid_EP3_result} (a). \cref{fig:solid-solid_EP3_result} (c) plots the trajectories of the three poles. The three trajectories form a single closed curve, which is a well-known characteristic of third-order EPs. Another important property is the asymptotic behavior of the degenerate poles when the parameters are perturbed at the third-order EP. We fix the parameter $\rho^{(2)}$ at the EP and perturb the other parameter as $R^{(2)} = 0.38351 R + \Delta_R$, where $\Delta_R$ is sufficiently small. The three poles should deviate from the degenerate value as the value of $\Delta_R$ increases. We define the variation as $\Delta_\omega = \mathrm{max}_{i\neq j} |\omega_i - \omega_j|$, which is a function of $\Delta_R$, and plot its values in \cref{fig:solid-solid_EP3_distance}. The result shows that the variation asymptotically behaves as $O(\Delta_R^{1/3})$, while the second-order EP exhibits the square-root behavior. This is another characteristic of third-order EPs.

\subsection{Fluid background}
When the layered solid is immersed in a fluid medium, the system loses its energy due to the acoustic wave radiation. Here, we use the same optimization approach to show that the solid--fluid coupled systems exhibit EPs.

In this subsection, we fix the mass density $\rho>0$, bulk modulus $\kappa>0$, and outer radius $R^{(N)}=:R$. We first assume $N=2$ and compute the scattering poles for $\rho^{(1)} = 16\rho$, $\rho^{(2)}=4\rho$, $\mu^{(1)}=\mu^{(2)}=\kappa$, and $R^{(1)} = R/2$. Subsequenctly, we chose two scattering poles $\omega R/c$ at $6.4436 - 0.029230\mathrm{i}$ and $7.0062 - 0.035093\mathrm{i}$ and optimize the parameters $\rho^{(1)}$ and $R^{(1)}$.
% w1 = 6.4436116523142255 -0.029229682873249115j
% w2 = 7.006176485366477 -0.035093495554679686j

%
% solid-fluid
%
\begin{figure}
  \includegraphics[scale=0.2]{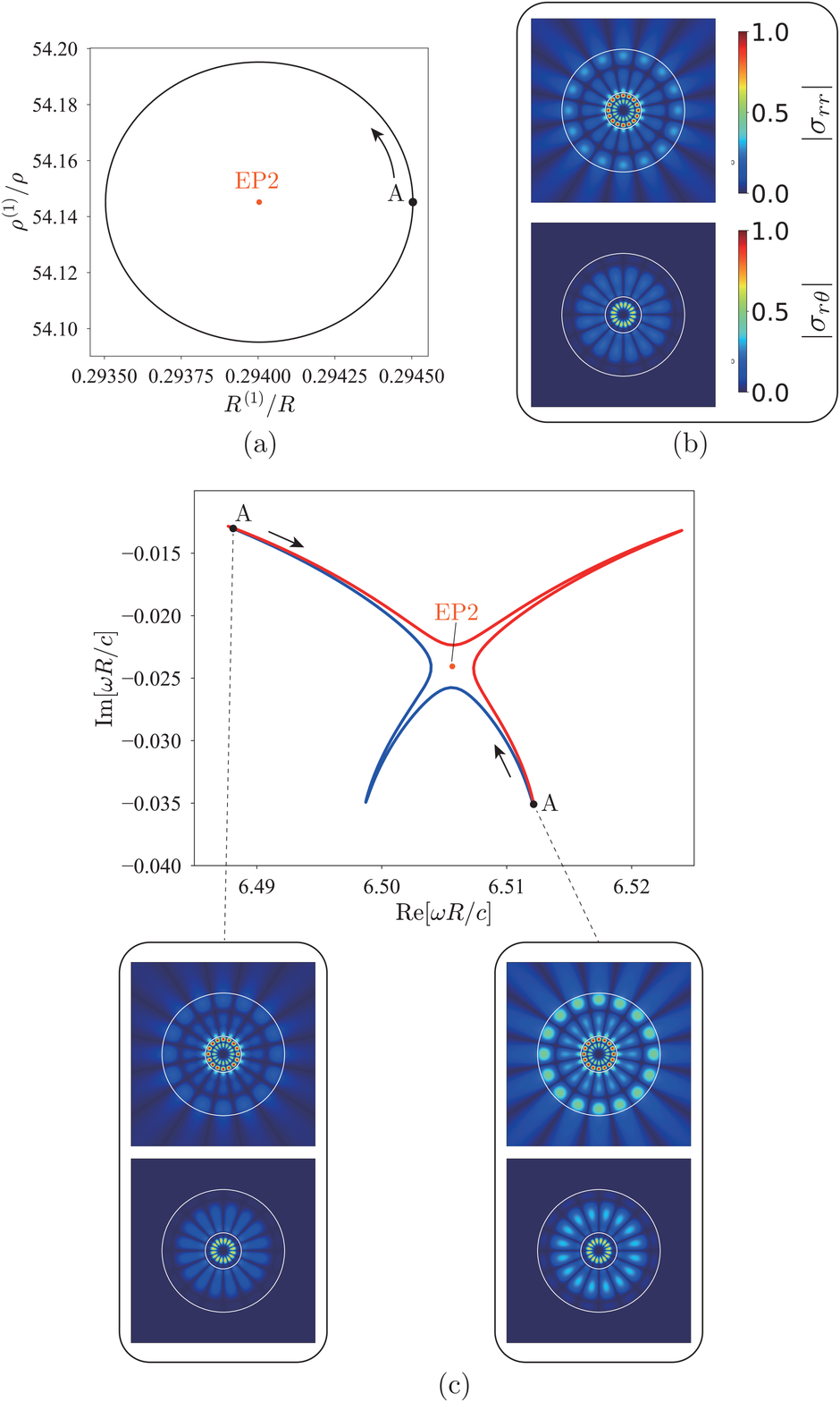}% Here is how to import EPS art
  \caption{\label{fig:solid-fluid_EP2_result} Encircling a second-order EP (EP2) in the solid--fluid system. (a) Encircling path in the parameter space. (b) Degenerate resonant mode at the EP2. (c) Trajectory of the two scattering poles. Each color corresponds to the single encircling shown in (a).}
\end{figure}
\cref{fig:solid-fluid_EP2_result} shows the results of the optimization and EP encircling. The parameter optimization successfully minimizes the objective function $J$ with two poles $\omega R/c$ at $6.5056469 - 0.0240559\mathrm{i}$ and $6.5056469 - 0.0240557\mathrm{i}$. The degenerate resonant modes are plotted in \cref{fig:solid-fluid_EP2_result} (b). Encircling the obtained parameters $R^{(1)}/R = 0.29400$ and $\rho^{(1)}/\rho = 54.145$ on the $R^{(1)}$--$\rho^{(1)}$ plane as shown in \cref{fig:solid-fluid_EP2_result} (a), we obtain the trajectories of the two poles on the complex $\omega$-plane (\cref{fig:solid-fluid_EP2_result} (c)) with the same structures as the second-order EP of the solid--solid model. From the results, we conclude that a second-order EP exists within the parameter path.
% 6.50564685969782e+00_-2.40558630633677e-02.dat  6.50564686372405e+00_-2.40556717972379e-02.dat

\begin{figure}
  \includegraphics[scale=0.2]{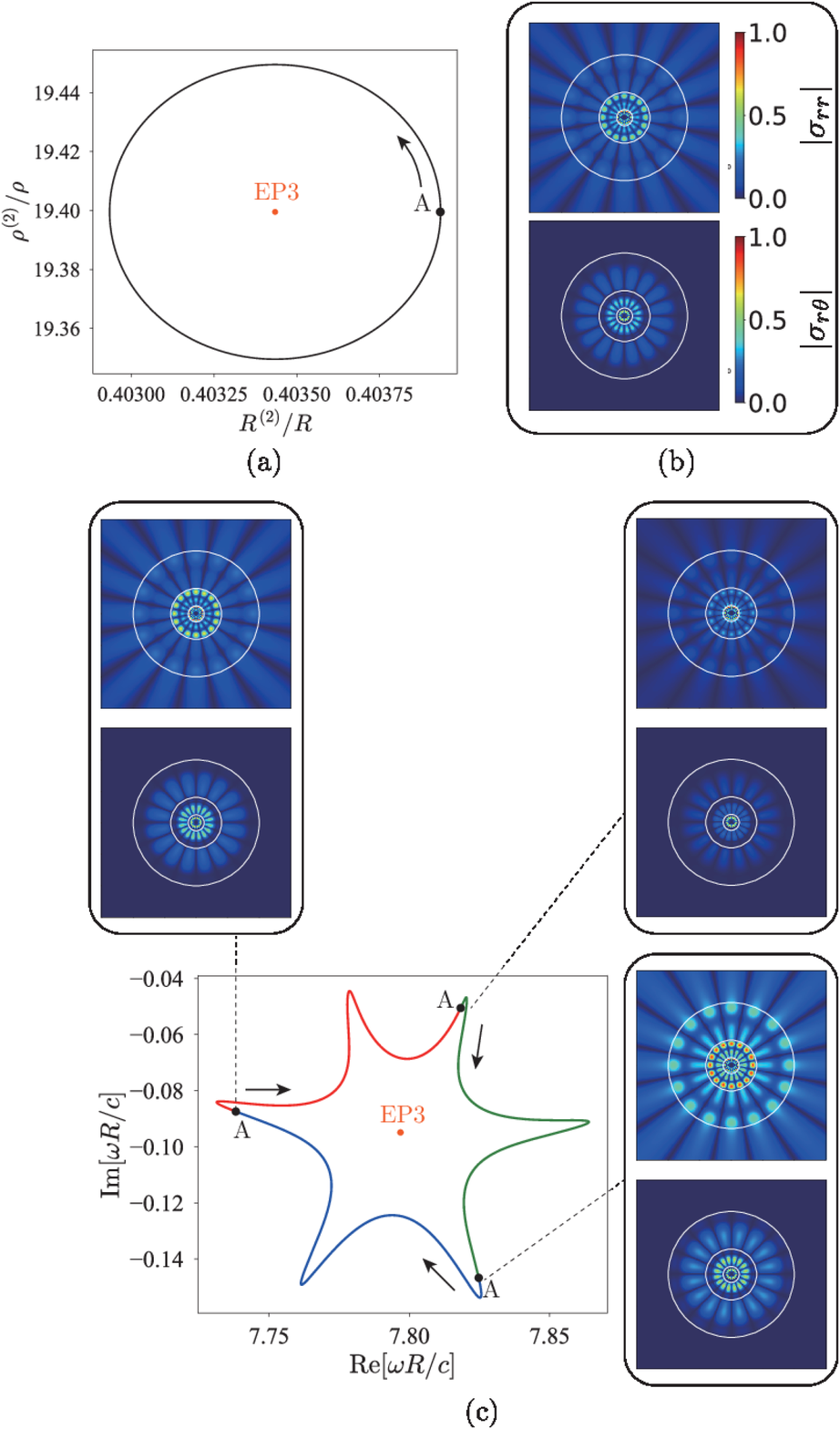}% Here is how to import EPS art
  \caption{\label{fig:solid-fluid_EP3_result} Encircling a third-order EP (EP3) in the solid-fluid system. (a) Encircling path in the parameter space. (b) Degenerate resonant mode at the EP3. (c) Trajectory of the two scattering poles. Each color corresponds to a single encircling shown in (a).}
\end{figure}
Finally, we shall prove the existence of a third-order EP. As with the solid--solid model, we use the triple-layered solid ($N=3$) with the following parameters:
\begin{itemize}
  \item Mass density $\rho^{(1)}=64\rho,\,\rho^{(2)}=16\rho,\,\rho^{(3)}=4\rho$
  \item Shear modulus $\mu^{(1)}=\mu^{(2)}=\mu^{(3)}=\kappa$
  \item Radii $R^{(1)} = R/4$, $R^{(2)} = R/2$, $R^{(3)} =: R$.
\end{itemize}
The system has three scattering poles $\omega R/c$ at $6.4222 - 0.022268\mathrm{i}$, $6.7600 - 0.025502\mathrm{i}$, and $7.2757-0.019004\mathrm{i}$. We optimize the parameters $R^{(1)}$, $R^{(2)}$, $\rho^{(1)}$, $\rho^{(2)}$, and $\rho^{(3)}$ with the other parameters fixed. 
%w1 = 6.42219559570099 -0.02226808460654044j
%w2 = 6.760033474140678 -0.02550155450959887j
%w3 = 7.2756856813508834 -0.01900374869021539j
\cref{fig:solid-fluid_EP3_result} illustrates the results. The three poles almost coalesce at $R^{(1)}/R=0.12678$, $R^{(2)}=0.40343$, $\rho^{(1)}/\rho = 200.46$, $\rho^{(2)}/\rho=19.400$, and $\rho^{(3)}/\rho=2.7136$ with $\omega_1 R/c = 7.7966-0.094927\mathrm{i}$, $\omega_2 R/c = 7.7971-0.094901\mathrm{i}$, and $\omega_3 R/c = 7.7968-0.094903\mathrm{i}$. Their resonant modes are also degenerated as shown in \cref{fig:solid-fluid_EP3_result} (b). The existence of a third-order EP is confirmed by the EP encircling. \cref{fig:solid-fluid_EP3_result} (a) and (c) present the results of the EP encircling. Like the solid--solid case (\cref{fig:solid-solid_EP3_result}), we observe a three-fold structure on the complex $\omega$-plane, which implies the existence of a three-order EP.
% 7.79656954-0.09492671j 7.79705694-0.09490135j 7.79681844-0.09490256j

\section{Conclusions}
This study discussed the non-Hermitian degeneracy of the scattering poles in cylindrical elastic systems with radiation loss. We first formulated two-dimensional elastodynamic problems using Lam\'e potentials and cylindrical functions. Some system parameters were optimized such that two or three scattering poles coalesce on the complex frequency plane. We performed the EP encircling around the optimized parameters to confirm the existence of second- and third-order EPs. We found that the sensitivity of degenerate poles at a third-order EP is $O(\Delta^{1/3})$, where $\Delta$ is a system parameter. This anomalous sensitivity allows enhancement on mechanical sensors, such as pressure/stress sensing and defect detection.

\begin{acknowledgments}
This work was supported by JSPS KAKENHI Grant Number JP22K14166.
\end{acknowledgments}

\appendix

\section{Explicit forms of the matrices}

To formulate Eq. \cref{eq:among_layers}, we use the continuity of $u$ and $\sigma\cdot n$, which is equivalent to that of $u_r$, $u_\theta$, $\sigma_{rr}$, and $\sigma_{r\theta}$. After some calculations \cite{eringen1975elastodynamics}, we have
\begin{align}
  M^{(l)}(r)
=
% W11 -> U1, W12 -> U2, W41 -> V1, W42 -> V2
\begin{bmatrix}
U^{(l)}_1(r) & U^{(l)}_2(r) & \tilde{U}^{(l)}_{1}(r) & \tilde{U}^{(l)}_{2}(r)
\\
V^{(l)}_1(r) & V^{(l)}_2(r) & \tilde{V}^{(l)}_{1}(r) & \tilde{V}^{(l)}_{2}(r)
\\
T^{(l)}_{11}(r) & T^{(l)}_{12}(r) & \tilde{T}^{(l)}_{11}(r) & \tilde{T}^{(l)}_{12}(r)
\\
T^{(l)}_{41}(r) & T^{(l)}_{42}(r) & \tilde{T}^{(l)}_{41}(r) & \tilde{T}^{(l)}_{42}(r)
\end{bmatrix}
,
\end{align}
where
\begin{align}
  T^{(l)}_{11} &= \frac{2\mu^{(l)}}{r^2}\biggl[ (m^2+m-\frac{1}{2}(k_\mathrm{T}^{(l)})^2r^2)J_m(k_\mathrm{L}^{(l)} r)
  \notag\\
  &\hspace{100pt} -k_\mathrm{L}^{(l)} rJ_{m-1}(k_\mathrm{L}^{(l)} r) \biggr],
  \\
  T^{(l)}_{12} &= \frac{2\mu^{(l)}}{r^2}\bigl[  -\mathrm{i} m \bigl( (m+1)J_m(k_\mathrm{T}^{(l)} r)
  \notag\\
  &\hspace{100pt} 
  -k_\mathrm{T}^{(l)} rJ_{m-1}(k_\mathrm{T}^{(l)} r) \bigr) \bigr],
  \\
  T^{(l)}_{41} &= \frac{2\mu^{(l)}}{r^2}\bigl[ -\mathrm{i} m \bigl( (m+1)J_m(k_\mathrm{L}^{(l)} r)
  \notag\\
  &\hspace{100pt} 
  -k_\mathrm{L}^{(l)} rJ_{m-1}(k_\mathrm{L}^{(l)} r) \bigr)  \bigr],
  \\
  T^{(l)}_{42} &= \frac{2\mu^{(l)}}{r^2}\biggl[ -(m^2+m-\frac{1}{2}(k_\mathrm{T}^{(l)})^2r^2)J_m(k_\mathrm{T}^{(l)} r)
  \notag\\
  &\hspace{100pt} 
  +k_\mathrm{T}^{(l)} rJ_{m-1}(k_\mathrm{T}^{(l)} r) \biggr],
  \\
  U^{(l)}_1 &= k_\mathrm{L}^{(l)} J^\prime_m (k_\mathrm{L}^{(l)} r),
  \\
  U^{(l)}_2 &= \frac{\mathrm{i}m}{r} J_m (k_\mathrm{T}^{(l)} r),
  \\
  V^{(l)}_1 &= \frac{\mathrm{i}m}{r} J_m (k_\mathrm{L}^{(l)} r),
  \\
  V^{(l)}_2 &= -k_\mathrm{T}^{(l)} J^\prime_m (k_\mathrm{T}^{(l)} r).
\end{align}
The functions with the tilde are defined by replacing $J_m$ with $H^{(1)}_m$.

Similarly, the matrix $Y_\mathrm{s}$ is given by
\begin{align}
  Y_\mathrm{s} = 
  \begin{bmatrix}
    \tilde{U}_{1}(R^{(N)}) & \tilde{U}_{2}(R^{(N)})
    \\
    \tilde{V}_{1}(R^{(N)}) & \tilde{V}_{2}(R^{(N)})
    \\
    \tilde{T}_{11}(R^{(N)}) & \tilde{T}_{12}(R^{(N)})
    \\
    \tilde{T}_{41}(R^{(N)}) & \tilde{T}_{42}(R^{(N)})
    \end{bmatrix}
    .
\end{align}

The solid--fluid coupling matrices are written as
\begin{align}
  Z_\mathrm{f} &= 
  \begin{bmatrix}
    T^{(l)}_{11}(R^{(N)}) & T^{(l)}_{12}(R^{(N)}) & \tilde{T}^{(l)}_{11}(R^{(N)}) & \tilde{T}^{(l)}_{12}(R^{(N)})
    \\
    T^{(l)}_{41}(R^{(N)}) & T^{(l)}_{42}(R^{(N)}) & \tilde{T}^{(l)}_{41}(R^{(N)}) & \tilde{T}^{(l)}_{42}(R^{(N)})
    \\
    U^{(l)}_1(R^{(N)}) & U^{(l)}_2(R^{(N)}) & \tilde{U}^{(l)}_{1}(R^{(N)}) & \tilde{U}^{(l)}_{2}(R^{(N)})
    \end{bmatrix}
    ,\\
    Y_\mathrm{f} &=
    \begin{bmatrix}
      -H^{(1)}_n (\omega R^{(N)}/c)
      \\
      0
      \\
      \frac{1}{c\rho\omega} H^{(1)\prime}_n (\omega R^{(N)}/c)
    \end{bmatrix}
    .
\end{align}

%\nocite{*}

%\bibliography{export}

\end{document}